\documentclass{amsart}
\usepackage{amsfonts}
\usepackage{amsmath}
\usepackage{amssymb}
\usepackage{amsthm}

\theoremstyle{theorem}
\newtheorem{theorem}{Theorem}

\begin{document}

\title{Existence of a simple and equitable fair division: a short proof
}
\markright{Equitable and simple fair division}
\author[G.~Ch\`eze]{Guillaume Ch\`eze}
\address{Guillaume Ch\`eze: Institut de Math\'ematiques de Toulouse\\
Universit\'e Paul Sabatier Toulouse 3 \\
MIP B\^at. 1R3\\
31 062 TOULOUSE cedex 9, France}
\email{guillaume.cheze@math.univ-toulouse.fr}

\maketitle

\begin{abstract}
In this note we study how to share a good between $n$ players in a simple and equitable way.  We give a short proof for the existence of such fair divisions. \\
\end{abstract}

 In this note we study a fair division problem. Fair divisions problems are sometimes called cake cutting problems. These kinds of problems appear when we study division of land, time or computer memory between different agents with different point of view.  This problem is old: the  ``cut and choose" algorithm already appears in the Bible. In a more scientific way it has been formulated by Steinhaus in 1948, see \cite{Steinhaus}. Nowadays, there exists several articles, see e.g. \cite{DubinsSpanier,BramsTaylorarticle,RoberstonWebbarticle,BJK}, and books, see e.g. \cite{RobertsonWebb,BramsTaylor, Procacciachapter,Barbanel}, about this topic. These results appears in the mathematics, economics, political science, artificial intelligence and computer science literature.\\
 

In this note, our heterogeneous good, e.g. a cake, will be represented by the interval $[0;1]$. 
We consider $n$ players and we associate to each player a non-atomic probability measure $\mu_i$ on the interval $X=[0;1]$ with density $f_i$. These measures represent the utility functions of the player. The set $X$ represents the cake and we want to get a partition of $X=X_1\sqcup \ldots \sqcup X_n$, where the $i$-th player get $X_i$.\\

In this situation several notions of fair division exists:
 \begin{itemize}
 \item Proportional  division: $\forall i,\, \mu_i(X_i) \geq 1/n$.
 \item Exact  division: $\forall i,\, \forall j,\, \mu_i(X_j)=1/n$.
 \item Envy-free division: $\forall i,\, \forall j,\, \mu_i(X_i)\geq \mu_i(X_j)$.
 \item Equitable division: $\forall i,\, \forall j,\, \mu_i(X_i) = \mu_j(X_j)$.\\
 \end{itemize}

  All these fair divisions are possible, see e.g \cite{Steinhaus,DubinsSpanier,BramsTaylor,RobertsonWebb,Cech}. The minimal number of cuts in order to get a fair division has also been studied.\\
  For a proportional fair division $n-1$ cuts are sufficient. This means that for all $i$ there exists an index $j$ such that $X_i= [x_{j},x_{j+1}]$, where $x_0=0$, $x_n=1$. A fair division where each $X_i$ is an interval is called a \emph{simple fair division}. The Banach-Knaster algorithm given in \cite{Steinhaus} shows how to get a simple proportional fair division.\\
   Stromquist and Woodall have shown that we can always get a simple envy-free fair division, see \cite{Stromquist, Stromquist2,Woodall}. These two different proofs use the Brouwer fixed point theorem. \\
  For an exact division $n(n-1)$ cuts are sufficient and this bound is also optimal, see \cite{Alon}. This means that we cannot always get an exact division where each $X_i$ is an interval. 
    The strategy used by Alon to prove his result relies on a general version of the Borsuk-Ulam theorem. \\
    The Borsuk-Ulam theorem is a classical tool in fair division. Indeed, it implies the Ham sandwich theorem see \cite{DubinsSpanier}, and it is also used to get a consensus halving, see e.g. \cite{DubinsSpanier,SimmonsSu}. Consensus halving means that we want to get two sets $A$ and $B$ such that $[0,1]=A \sqcup B$ and for all $i$, $\mu_i(A)=\mu_i(B)$. This result corresponds to the Hobby-Rice theorem and a proof using the Borsuk-Ulam theorem has also been given by Pinkus, see \cite{Pinkus}. As Borsuk-Ulam theorem implies the Brouwer fixed point theorem we can say that the existence of a simple and envy-free fair division relies on the Borsuk-Ulam theorem. This suggests  that the Borsuk-Ulam theorem  is an ubiquitous tool in cake-cutting problems. \\
However  existing proofs about equitable and simple fair division do not use this theorem. Proofs have been given by Cechl\'arov\'a, Dobo\v{s} and Pill\'arov\'a in \cite{Cech} and independently by Aumann and Dombb in \cite{Aumann}.
In \cite{Cech} the authors  have proved that for all permutations $\sigma \in \mathcal{S}_n$ the system of equations:
$$(\star) \quad \int_0^{x_1} f_{\sigma(1)}(x) dx=\int_{x_1}^{x_2} f_{\sigma(2)}(x)dx=\cdots=\int_{x_{n-1}}^1 f_{\sigma(n)}(x)dx,$$
has a solution with $0 \leq x_1 \leq x_2 \leq \cdots \leq x_{n-1} \leq 1$. The permutation $\sigma$ corresponds to the players' order: the $\sigma(i)$-th player get the interval $[x_{i-1},x_i]$.\\
The proof given in \cite{Cech} is based on the notion of the  generalized inverse of a function. 
In \cite{Aumann} the authors give another proof of this result which uses a compactness argument. \\
Thus we have two different proofs but these proofs do not use the classical tool: the Borsuk-Ulam theorem. \\
In the following we show how to get the existence of an equitable and simple fair division  shortly thanks to this classical theorem.

\begin{theorem}
For all densities function $f_i$ and all permutations $\sigma \in \mathcal{S}_n$ the system of equations $(\star)$ has a solution. This means that there exists an equitable and simple fair division. 
\end{theorem}

Our proof relies on the Borsuk-Ulam theorem: If $f:S^k\rightarrow \mathbb{R}^k$ is  continuous and antipodal ($f(-x)=-f(x)$) then there exists $x_0 \in S^k$ such that $f(x_0)=0$.

\begin{proof}
We consider the sphere $S^{n-1}=\{e=(e_1,\dots,e_n) \in \mathbb{R}^{n} | \sum_{i=1}^n e_i^2=1 \}$ and the function
\begin{eqnarray*}
f:S^{n-1} & \rightarrow & \mathbb{R}^{n-1}\\
e&\longmapsto& \big(F_1(e),\ldots,F_{n-1}(e)\big)
\end{eqnarray*}
where 
$$F_i(e)= sgn(e_{i+1}).\int_{e_1^2+\cdots+e_{i}^2}^{e_1^2+\cdots+e_{i}^2+e_{i+1}^2} f_{\sigma(i+1)}(x) dx  - sgn(e_1).\int_{0}^{e_1^2} f_{\sigma(1)}(x) dx.$$
The function $f$ is continuous and antipodal, thus by the Borsuk-Ulam theorem there exists $\tilde{e} \in S$ such that $f(\tilde{e})=0$.\\
We set $x_0=0$ and $x_i=x_{i-1}+\tilde{e}_i^2$ and we get the desired solution for the system $(\star)$.

\end{proof}

\bibliographystyle{alpha} 
 
\bibliography{cakebiblio}

\end{document}